\documentclass{conm-p-l}

\usepackage{amsthm}
\usepackage{amsmath}
\usepackage[initials]{amsrefs}
\usepackage{graphicx}

\theoremstyle{definition}
\newtheorem{example}{Example}

\newcommand\D{{\bf D}}
\renewcommand\S{{\bf S}}
\newcommand\U{{\bf U}}
\newcommand\M{{\bf M}}
\newcommand\p{{\bf p}}
\newcommand\q{{\bf q}}
\newcommand\vv{{\bf v}}
\newcommand\cB{\mathcal{B}}
\newcommand\cP{\mathcal{P}}
\newcommand\cV{\mathcal{V}}
\newcommand\bbar[1]{\overline{#1}}
\newcommand\e{{\rm e}}

\begin{document}

\title[Form factor expansion for large graphs]{Form factor expansion
  for large graphs: a diagrammatic approach} 

\author{Gregory Berkolaiko}

\address{Department of Mathematics, Texas A\&M University, College
  Station, TX 77843-3368, USA}

\subjclass[2000]{81Q50,34B45,15A52}

\begin{abstract}
  The form factor of a quantum graph is a function measuring
  correlations within the spectrum of the graph.  It can be expressed
  as a double sum over the periodic orbits on the graph.  We propose a
  scheme which allows one to evaluate the periodic orbit sum for a
  special family of graphs and thus to recover the expression for the
  form factor predicted by the Random Matrix Theory.  The scheme,
  although producing the expected answer, undercounts orbits of a
  certain structure, raising doubts about an analogous summation
  recently proposed for quantum billiards.
\end{abstract}

\maketitle

\section{Introduction}

One of the central questions of quantum chaology is investigating
which properties of the spectrum of a quantum system reflect the
chaoticity of the underlying dynamics.  It has been observed
\cite{BGS84} that if the classical limit is chaotic, the local
statistical properties of the spectrum of a {\em generic} quantum
system resemble those of the spectrum of a large matrix from a
suitable random matrix ensemble.  The choice of the ensemble depends
only on very general characteristics of the system, such as whether it is
time-reversal invariant, and whether spin is present.  Clarifying the
precise meaning of the term ``generic'' above and proving the
conjecture of \cite{BGS84} in any generality still remains one of the
most exciting questions of quantum chaology.

Quantum graphs make ideal models to study the questions of quantum
chaology, as they exhibit the same rich variety of behaviors and share
the same set of analytical tools as the more complicated systems.
However one can achieve deeper understanding on graphs while using
methods which are simpler yet more rigorous.  And due to direct
analogies between graphs and more complicated systems like billiards
and cavities, any advancement on graphs will provide insight to
researchers in a much wider area.  It is notable that some of the
recent progress in understanding the role of periodic orbits in the
spectral statistics was achieved on graphs first \cites{BSW03,B04} and
then extended to other systems \cite{HMBH04}.

In the present paper we aim to exploit the analogy between the tools
used in quantum chaology on billiards and graphs to analyze a recent
breakthrough result by M\"uller {\em et al\/} \cite{MHBHA05}.  The
authors of \cite{MHBHA05} investigate the form factor, one of the
functions measuring the correlations within the spectrum of a
classically chaotic quantum billiard.  The advantage of the form
factor lies in its convenient expansion in terms of pairs of periodic
orbits of the classical system.  This expansion is an application of the
trace formulae (see \cite{CPS98} for a review) and has been a starting
point for many investigations.  The progress in understanding the role
of periodic orbits in the universal behavior of the form factor of
billiard systems is marked by such milestones as the diagonal
approximation \cite{Ber85}, the first off-diagonal contribution
\cites{Sie02,SR01} and, most recently, \cite{MHBHA05} which announced a
complete expansion of the form factor for the rescaled time $\tau<1$.
For quantum graphs, the corresponding steps have been taken in
\cites{KS97,KS99,Tan01} (the diagonal approximation) and \cite{BSW02}
(the first off-diagonal contribution).  This article aims to lay the
groundwork needed for the complete expansion.  It must be mentioned
that the form-factor for graphs has also been successfully analyzed
using non-perturbative supersymmetric methods \cites{GA04,GA05}.

To introduce our results we need to briefly summarize the achievements
of \cite{MHBHA05}.  Using an expansion of the form factor as a double
sum over the periodic orbits of the classical system the authors
classify the periodic orbit pairs and perform the summation to recover,
for $\tau<1$, the result predicted by the Random Matrix Theory.  The
summation can be loosely divided into two parts: evaluating the
contribution of pairs of orbits related through a fixed transformation
(described as a ``diagram'') and summation of the resulting
contributions over all possible transformations.  While performing
breathtaking (and mathematically rigorous) feats of combinatorics in
the second part, the authors remain somewhat vague on the assumptions
made about the ergodic nature of the periodic orbits in the first
part.  By analyzing the contributions of the diagrams on quantum
graphs we aim to look for possible omissions in the derivation of
\cite{MHBHA05}.

We are pleased to report that, in the large, we are able to replicate
and thus confirm the findings of \cite{MHBHA05}.  In doing so,
however, we are forced to omit a class of periodic orbits that do not
fit well into the general summation.  We believe that such orbits were
also omitted (although less explicitly) in \cite{MHBHA05}.  These are
the orbits with short stretches between the self-intersections.  On
one hand, if the conjecture of \cite{BGS84} were to hold, such orbits
should not contribute to the universal final answer: the short
stretches are {\em not} universal.  For exactly the same reason the
task of dealing with such orbits is one of the hardest.  A general
method of proving that their contribution is negligible is badly
needed for higher terms.

The paper is organized as follows: in section~\ref{sec:setup} we
discuss the definition of the quantum graphs, their spectra, the
relevant statistics, the trace formula and the particular model
considered.  Then we review the past results and difficulties
encountered in section~\ref{sec:past} and proceed to describe our new
summation method in section~\ref{sec:newmethod}.  Our results are
compared with those of M\"uller {\em et al\/} \cite{MHBHA05} in
section~\ref{sec:conclusion}, where we also speculate about the
possibility of periodic orbit summation beyond the Heisenberg time
$\tau=1$.

\section{Spectral statistics of quantum graphs}
\label{sec:setup}

\subsection{Quantum dynamics on graphs}
\label{sec:graphs}

Here we will describe a general approach to quantization of a graph
proposed by Schanz and Smilansky \cite{SS99}.  It shares many
important features with another quantization procedure, used by
several authors in the present volume (see, e.g.,
\cite{kostrykin_schrader}), which is based on finding a self-adjoint
extension of a second-order differential operator defined on the bonds
of a graph.  In particular, in many important cases (such as the
Neumann boundary conditions) the resulting quantum spectra coincide.

We start with a graph $G = (\cV,\cB)$ where $\cV$ is a finite set of
vertices (or nodes) and $\cB$ is the set of directed bonds (or edges).
Each bond has a start vertex and an end vertex, they are specified via
functions $s:\cB\to\cV$ and $e:\cB\to\cV$.  The start and end vertices
specify a bond uniquely: we do not allow multiple edges.  We will,
however, consider graphs with looping bonds, i.e.~$b$ such that
$s(b)=e(b)$.  We demand that the set $\cB$ be symmetric in the sense
that $b \in \cB$ iff there is a bond $\bbar{b} \in \cB$ such that
$s(b)=e(\bbar{b})$ and $e(b)=s(\bbar{b})$.  The bond $\bbar{b}$ is
called the {\em reversal} of $b$.  A looping bond acts as its own
reversal.  The operation of reversal is reflexive: $\bbar{\bbar{b}} =
b$.  When we refer to the ``non-directed bond $b$'' we mean the couple
of bonds, $b$ and $\bbar{b}$.  The number of vertices is denoted by $V
= |\cV|$ and the number of directed bonds is $B = |\cB|$.  The
vertices are usually marked by the integers starting from 0.

The graphs we will be considering are metric, that is each bond $b$
has a length denoted by $L_b$.  Naturally, $L_b = L_{\bbar{b}}$.  As a
rule, we will be assuming that the different lengths are rationally
independent, which means that the only choice of integers
$\{k_b\}_{b\in\cB}$ satisfying
\begin{equation}
  \sum_{b\in\cB} k_b L_b = 0
\end{equation}
is the trivial one: $k_b=0$ for all $b\in\cB$.

Now we define the quantum scattering map to be a $B\times B$ unitary
matrix
\begin{equation}
  \label{eq:quantum_prop}
  \U(k) = \D(k)\S.
\end{equation}
Here the matrix $\D$ is diagonal with the elements $e^{ikL_b}$, where $L_b$
is length of the bond $b$.  An element $S_{b_2,b_1}$ of the unitary
matrix $\S$ is zero unless the transition between $b_1$ and $b_2$ is
possible according to the graph's geometry (i.e.~$e(b_1)=s(b_2)$).
The evolution is defined by $\psi_{n+1} = U(k) \psi_n$, where $\psi$
are $B$-dimensional vectors of the wave amplitudes on the graph's
bonds.  The eigenvalues of the system are the wavenumbers $k$ at which
stationary solutions exist, i.e.~$\U(k)$ has an eigenvalue one.  The
quantum evolution is time-reversal (TR) invariant if the elements of
the matrix $\S$ satisfy
\begin{equation}
  \label{eq:tr_inv_S}
  S_{b_2,b_1}=S_{\bbar{b_1},\bbar{b_2}}.
\end{equation}

The corresponding classical dynamics on the graph is defined as the
Markov chain on the bonds of the graph with the transition matrix $\M^{(B)}$,
\begin{equation}
  \label{eq:trans_matr}
  M_{b_2,b_1} = |U_{b_2,b_1}|^2 = |S_{b_2,b_1}|^2.
\end{equation}
One of the most exciting questions of quantum chaology on graphs is to
relate the statistical properties of the spectrum of the quantum graph
to the ergodic properties of the corresponding Markov chain.

\subsection{Considered model}
\label{sec:model}

In this paper we will only consider a particular family of graphs,
complete Fourier graphs.  In a complete graph, there is a bond between
any two vertices, including a loop from each vertex to itself.  In a
graph with $N$ vertices, $B$ is equal to $N^2$.  The corresponding $\S$
matrix is defined by
\begin{equation}
  \label{eq:fourier}
  S_{b_2,b_1} = \delta_{e(b_1),s(b_2)}
  \frac{1}{\sqrt{N}} e^{jm\frac{2\pi i}N},
\end{equation}
where $j=s(b_1)$ and $m=e(b_2)$ are integers between 1 and $N$.  The
distinguishing feature of a complete Fourier graph is the quick
equilibration of the corresponding classical transition probability:
the probability to go from a bond $b_1$ to a bond $b_2$ in $t$ steps
is equal to $1/B$ independently of $b_1$ and $b_2$ as soon as $t\geq
2$.

\subsection{Spectral statistics and the BGS conjecture}
\label{sec:stats}

Consider an ordered sequence $\{k_i\}_{i=1}^\infty$ with mean density
\begin{equation}
  \label{eq:mean_dens}
  \bbar{d} = \lim_{N\to\infty} \frac{\#\{j : k_j < N\}}{N}
\end{equation}
equal to 1.  The most popular quantity for numerical study of the statistical
properties of the sequence $\{k_i\}_{i=1}^\infty$ is the nearest-neighbor
spacing distribution, which is the weak limit (if it exists)
\begin{equation}
  \label{eq:nnsd}
  p(x) = \lim_{N\to\infty} \frac1N \sum_{i=1}^N \delta(x-(k_{i+1}-k_i)).
\end{equation}
The two-point correlation function is a second order analogue of
$p(x)$ and is defined by
\begin{equation}
  \label{eq:R_2}
  R_2(x) = \lim_{N\to\infty} \frac1N \sum_{i=1}^N\sum_{j=1}^N 
  \delta(x-(k_j-k_i)).
\end{equation}
It is important to note that the normalization in front of the sum is
$N^{-1}$ and not $N^{-2}$, i.e. $R_2(x)$ is not a probability
distribution.  The most popular quantity for analytical investigations
is the Fourier transform of $R_2(x)$, the form factor,
\begin{equation}
  \label{eq:ff_defn}
  K(\tau) = \int_{-\infty}^\infty e^{2\pi i x\tau} \left(R_2(x)-1\right) dx.
\end{equation}
It owes much of its popularity to its convenient representations via
trace formulae.

\begin{example}
  {\bf Uncorrelated spectrum.}  Let $\{\xi_i\}_{i=1}^\infty$ be a sequence of
  independent exponentially distributed random variables with expectation 1.
  Define $k_i = \sum_{j=1}^i \xi_j$, these are the jump times of a Poisson
  process \cite{Billingsley}.  One can show that $R_2(x)=1+\delta(x)$ for the
  sequence $\{k_i\}$, i.e. there are no detectable correlations within this
  sequence.
\end{example}

The Bohigas-Giannoni-Schmit (BGS) conjecture \cite{BGS84}, if applied
to quantum graphs, claims that the spectrum of quantum graphs is
correlated in the same way as in the spectrum of large random matrices
\cites{Mehta,Haake}.  In the case of TR invariant quantum graphs, the
random matrix should be from Gaussian Orthogonal Ensemble (GOE) or
Circular Orthogonal Ensemble\footnote{Loosely speaking, GOE
  corresponds to the spectrum of the Hamiltonian whereas COE
  corresponds to the spectrum of the $S$-matrix.  The spectral
  statistics are the same.} (COE).  In particular, the form factor of
graphs, $K^{(B)}(\tau)$ should converge to the form factor for the
Gaussian Orthogonal Ensemble (GOE),
\begin{equation}
  \label{eq:K_GOE}
    K(\tau) = 
  \begin{cases}
    2\tau - \tau\ln(1+2\tau), & \tau \leq 1,\\
    2 - \tau\ln\left(\frac{2\tau+1}{2\tau-1}\right), & \tau \geq 1,
  \end{cases}
\end{equation}
as we increase the number of bonds $B$.  Since often there is no
unique way of increasing the size of the graphs, a separate conjecture
describes the sequences that are expected to produce convergence to
the RMT statistics.  This conjecture, often called the ``Tanner
conjecture'' was formulated in \cite{Tan01} (see also \cites{GA04,GA05}); a
related criterion was proposed in \cites{BSW02,BSW03}.  They can be
summarized as requiring the mixing time of the Markov chain $\M^{(B)}$
to grow sufficiently slowly with $B$.  Our model, the complete Fourier
graphs, clearly satisfies this requirement: the mixing time is
independent of $B$.

\subsection{Trace formula}
\label{sec:trace}

One of the main links connecting a quantum system to the underlying
dynamics is a trace formula which expresses the eigenvalues of the
quantum system through a sum over the periodic orbits of the
classical system.  For quantum graphs the trace formula was established
by Roth in \cites{Roth83a,Roth83b} and rediscovered by Kottos and Smilansky in
\cite{KS97}.  Before we can write it down, we need to give several
definitions.  Let $\widetilde{\cP}_n$ be the set of all sequences
\begin{equation}
\label{eq:seq_compat}
  \p = [b_1, b_2, \ldots, b_n], \qquad b_i\in \cB, \qquad
  n\geq2 
\end{equation}
such that for any $i=1,\ldots,n$ the bond $b_{i+1}$ follows the bond
$b_i$: $e(b_i)=s(b_{i+1})$ (by $b_{n+1}$ we understand $b_1$).  Define
the cyclic shift operator $\sigma$ on $\widetilde{\cP}_n$ by
\begin{equation}
  \sigma\Big( [b_1, b_2, \ldots, b_n] \Big) = [b_2, b_3,
  \ldots, b_n, b_1].
\end{equation}
Equivalence classes in $\widetilde{\cP}_n$ with respect to the shift
$\sigma$ are called {\it periodic orbits} of period $n$.  For a
periodic orbit $\p=(b_1,\ldots,b_n)$ we define its length and
amplitude by
\begin{equation}
  \label{eq:length_ampl}
  l_\p = \sum_{i=1}^n L_{b_i} 
  \qquad\mbox{and}\qquad 
  A_\p = \prod_{i=1}^n S_{b_{i+1},b_i}
\end{equation}
correspondingly.  A periodic orbit $\p$ is a {\it repetition\/} if it
can be represented as another orbit repeated $r$ times.  The maximum
of all such $r$ is called the {\it repetition number\/} and denoted $r_\p$.
An orbit with $r_\p=1$ is called {\it prime}.  If $n$ is a prime
number, all orbits of period $n$ are prime.

The {\em trace formula} establishes a connection between the set of all
eigenvalues $\{k_n\}_{n=1}^\infty$ and the set of all periodic orbits $\cP$,
\begin{equation}
  \label{eq:trace_form}
  d(k) \equiv \sum_n \delta(k-k_n) = \frac{L}{2\pi} + \frac1\pi
  \sum_{\p\in\cP} \frac{l_\p}{r_\p} A_\p \cos(kl_\p),
\end{equation}
where $L$ is the sum of all bond lengths of the graph.  The right-hand
side of (\ref{eq:trace_form}) is convergent in the sense of
distributions.  The trace formula is exact on quantum graphs and can
be used to recover individual eigenvalues \cite{BDJ02} as well as the
whole density of states.

\subsection{Eigenvalue vs eigenphase statistics}
\label{sec:eigenphase}

The trace formula, (\ref{eq:trace_form}), provides a convenient
expression for the form factor of eigenvalues $\{k_n\}_{n=1}^\infty$,
\begin{equation}
  \label{eq:K_tau}
  K_{S}(\tau) = \frac1{L^2} \sum_{\p,\q\in\cP}
  \frac{L_\p}{r_\p}\frac{L_\q}{r_\q} A_\p A_\q
  \delta\left(\tau-\frac{L_\p}{L}\right) \delta_{L_\p,L_\q},
\end{equation}
where $\p$ (corresp.~$\q$) is a periodic orbit of length $L_\p$
($L_\q$) and with ``stability'' amplitude $A_\p$ ($A_\q$).  As signified by
the last Kronecker delta, two orbits $\p$ and $\q$ will contribute to
the double sum only if the lengths of the orbits coincide.

It is even more convenient to consider the form factor of eigenphases
of the graph's scattering map.  For a unitary operator, such as the
matrix $\U(k)$ defined in (\ref{eq:quantum_prop}), the form factor is
defined at integer times $n=0,1,\dots$ by
\begin{equation}
  \label{eq:ff_unitary}
  K_U^{(B)}(\tau)=\frac{1}{B}\langle|{\rm tr}(\U(k))^n|^{2}\rangle,
\end{equation}
where $\tau$ is the scaled time $\tau=n/B$.  In our case the averaging
is performed either with respect to the lengths of individual bonds or
with respect to $k$.  Both lengths and $k$ enter only via the matrix
$\D$ and averaging produces equivalent results.  Expanding the powers
of the trace we obtain
\begin{equation}\label{eq:trace_formula_for_trace}
  {\rm tr}(\U(k))^n = \sum_{\p}\frac{n}{r_\p}A_\p \e^{i kL_\p}.
\end{equation}
Substituting this into the definition of the form factor and performing
the averaging we arrive to
\begin{equation}
  \label{eq:ff_unitary_po_rep}
  K_U^{(B)}(\tau)=\frac{1}{B} \sum_{\p,\q} \frac{n^2}{r_\p r_\q} 
  A_\p A_\q^* \delta_{L_\p,L_\q}.
\end{equation}
It is an accepted fact\footnote{although the author is unaware of a
mathematical proof of such a statement, the reasons behind this
correspondence are discussed, among other sources, in
\cites{DS92,KS99,schanz}} that it is equivalent to study the spectral
statistics of the sequence $\{k_i\}$ of eigenvalues and the statistics
of the eigenphases of $\U(k)$.  Certainly, if we send all individual
bond lengths to $1$ in a fixed graph with real $\S$, it is clear from
expressions (\ref{eq:K_tau}) and (\ref{eq:ff_unitary_po_rep}) that
$K_{S}(\tau)$ will converge to $K_U^{(B)}(\tau)$ in the sense of
distributions on any compact interval.

We are interested in the limit of large graphs, $B\to\infty$, where
$K_U^{(B)}(\tau)$ is expected to converge to its limiting GOE form,
\begin{eqnarray}
  \label{eq:kgoe}
  K_{\rm GOE}(\tau)&=&2\tau-\tau\log(1+2\tau)\qquad(0\le \tau\le 1)
  \nonumber\\
  &=& 2\tau-2\tau^{2}+2\tau^{3}+O(\tau^{4})\,.
\end{eqnarray}
When taking this limit we can ignore the orbits that are repetitions
of other orbits, simplifying the formula for the form factor to
\begin{equation}
  \label{eq:ff_unitary_po}
  K_U^{(B)}(\tau)=\frac{n^2}{B} \sum_{\p,\q} A_\p A_\q^* 
  \delta_{L_\p,L_\q}.
\end{equation}
Equation (\ref{eq:ff_unitary_po}) is the starting point for our derivation.

\section{Past results}
\label{sec:past}

\subsection{Diagonal approximation}
\label{sec:diagonal}

To approach the task of evaluating the form factor one needs to
classify possible pairs of orbits of equal length.  The most simple
such pair is $\p$ and $\q=\p$.  Since we assumed that $L_b =
L_{\bbar{b}}$, we can also take $\q$ to be the reversal of $\p$:
$\q=(\bbar{b_n},\ldots,\bbar{b_1})$.  It turns out \cite{Ber85} that
such pairs provide the leading contributing term to the form factor,
called the diagonal approximation.  Evaluating it for graphs was
studied in detail in \cite{Tan01}.  Here we recap the derivation of
the diagonal approximation for our model, the complete Fourier graph,
for which the number of directed edges is $N^2$ and the modulus square
of amplitude of any orbit of period $n$ is $N^{-n}$.
\begin{equation}
  \label{eq:diag}
  \begin{split}
    K_{{\rm diag}} &= \frac{n^2}{N^2} 
    \sum_\p \left(A_\p A_\p^* + A_\p A_{\bbar{\p}}^*\right)
    = \frac{n^2}{N^2} \sum_\p \left(|A_\p|^2 + |A_\p|^2\right)\\
    &= \frac{n^2}{N^2} \sum_\p \frac2{N^n} = 2\frac{n}{N^2} = 2\tau.
  \end{split}
\end{equation}
Here we used the fact that there are asymptotically $N^n/n$ orbits
on a complete graph (we divide by $n$ to take into account the shifts
of the same orbit) and that the rescaled time $\tau$ is $n/B= n/N^2$.

\subsection{The difficulties in evaluating further corrections}
\label{sec:difficulties}

\begin{figure}[t]
  \centerline{\includegraphics{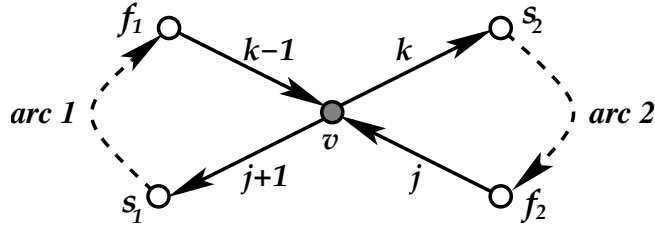}}
  \caption{Notation around a self-intersection.  The partner orbit is
    obtained by reversing the directions of the bonds number $k$ and
    $j$ and also reversing arc 2.}
  \label{fig:anatomy}
\end{figure}

To appreciate the difficulties present when evaluating off-diagonal
corrections, it is instructive to consider the simplest first
correction in some detail.  The first correction comes from the
Sieber-Richter pairs \cites{SR01,Sie02}, also called ``figure of
eight'' orbits.  We will say that an orbit $\p=(b_1,\ldots,b_n)$ has a
self-intersection at vertex $v$ if there are indices $k$ and $j$ such
that $k\neq j+1$ and $s(b_k)=e(b_j)=v$, see Fig.~\ref{fig:anatomy}.
Then the ``partner'' orbit
\begin{equation}
  \label{eq:partner_8}
  \q = (b_1,\ldots,b_{k-1}, \bbar{b_j}, \ldots, \bbar{b_k},
  b_{j+1},\ldots,b_n)
\end{equation}
will have the same length as $\p$ (and it is easy to verify that the
above $\q$ is a legitimate orbit).  We do
not require that $k<j$; the general rule is that we reverse all the
bonds starting with $b_k$ and until we have reversed $b_j$.

To evaluate the contribution of such pairs we need to introduce more
terminology, summarized in Fig.~\ref{fig:anatomy}.  We will usually
denote the intersection vertex by $v$.  The sequence of bonds
$[b_{j+2}, \ldots, b_{k-2}]$ we call arc 1; the sequence
$[b_{k+1}, \ldots, b_{j-1}]$ is arc 2.  The starting vertex of arc 1,
$s(b_{j+2})$ will be denoted $s_1$, the end vertex $e(b_{k-2})=f_1$.
Similarly for the second arc, $s(b_{k+1})=s_2$ and $e(b_{j-1})=f_2$.
We define $A_i$ to be the contribution to the amplitude made by
traversing arc $i$,
\begin{equation}
  \label{eq:A_arc1}
  A_1 = S_{b_{j+1},b_{j+2}}\cdots S_{b_{k-2},b_{k-1}}, \qquad 
  A_2 = S_{b_{k},b_{k+1}}\cdots S_{b_{j-1},b_{j}}.
\end{equation}
Using the above definition and the formula for $S_{b,b'}$ we can write
for $A_\p$ and $A_\q$
\begin{eqnarray}
  \label{eq:A_pq}
  A_\p &=& A_1 \times \frac1{\sqrt{N}}\e^{f_1s_2\frac{2\pi i}N} 
  \times A_2 \times \frac1{\sqrt{N}} \e^{f_2s_1\frac{2\pi i}N} \\
  A_\p &=& A_1 \times \frac1{\sqrt{N}}\e^{f_1f_2\frac{2\pi i}N} 
  \times A_{\bbar{2}} \times \frac1{\sqrt{N}} \e^{s_2s_1\frac{2\pi i}N},
\end{eqnarray}
where $A_{\bbar{2}}$ is defined by appropriately reversing the bonds
in $A_2$.  However, due to the reversal symmetry in the matrix $\S$,
$S_{b,b'}=S_{\bbar{b'},\bbar{b}}$, we have $A_{\bbar{2}}=A_2$.
Therefore, the contribution of the pair $(\p,\q)$ to the form
factor sum simplifies to
\begin{equation}
  \label{eq:pq_contrib}
  A_\p A_\q^* = |A_1|^2 |A_2|^2 \frac1{N^2} 
  \e^{(f_1s_2+f_2s_1 - f_1f_2-s_2s_1)\frac{2\pi i}{N}} 
  = \frac{1}{N^n} \e^{(f_2-s_2)(f_1-s_1)\frac{2\pi i}N},
\end{equation}
with the factorization in the exponent reminiscent of the expression
for the action difference in billiard systems \cites{Spe03,TR03}.

It is now clear how one can sum over all ``figure of eight'' orbits.
The sum will run over all possible intersection vertices, over
possible choices for $s_i$ and $f_i$ and over possible configurations
of arc 1 and 2 (including their length).  This approach, however, is
prone to counting orbits that should not be counted.  There are two
basic corrections.

\begin{figure}[t]
  \centerline{\includegraphics{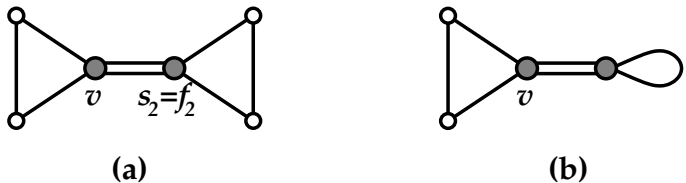}}
  \caption{Examples of orbits that would be miscounted in a naive
    summation.  In orbit (a) the intersection vertex is not uniquely
    defined, choosing either $v$ or $s_2$ would result in the same
    partner orbit.  In orbit (b) the partner orbit, obtained by
    reversal of the arc 2 coincides with the orbit itself, producing a
    pair which was counted in the diagonal approximation already.}
  \label{fig:corr}
\end{figure}

The first correction comes from the orbits in which $b_k=\bbar{b_j}$,
Fig.~\ref{fig:corr}~(a).  In this case the orbit runs along itself and
one can use $s_2$ as another intersection point.  It is clear,
however, that this choice would produce the same partner orbit $\q$.
A solution proposed in \cite{BSW02} was to define the intersection
point to be the leftmost vertex in such an ``extended'' intersection.
In essence, this amounts to only counting the intersections with
$s_1\neq f_1$.  This condition was called a {\em restriction}.  In the
present paper we modify this approach.

Another problem would arise if one of the arcs, say arc 2, was
self-retracing.  One of the simplest examples is when $s_2=f_2$ and
arc 2 contains just one bond, a loop from $s_2$ to itself,
Fig.~\ref{fig:corr}~(b).  In this case, reversing arc 2 would give
back the same orbit $\p$ and the pairing of $\p$ with itself was
already counted in the diagonal approximation.  Such orbits were
termed {\em exceptions} in \cite{BSW02}; they were evaluated
separately and subtracted from the previous result.  We note that the
previously imposed restriction $s_1\neq f_1$ means that there can be
no exceptions with arc 1 being self-retracing.  This shows that
restrictions and exceptions cannot be treated independently.  It
should also be noted that both restrictions and exceptions must be
treated carefully in order to get the correct result.

To find higher order terms, orbits fitting more complicated
intersection patterns have to be considered.  Diagrams were used in
\cite{BSW03} and \cite{B04} to describe the relationship between the
orbits $\p$ and $\q$ (i.e. which arcs get reversed).  It proved to be
hard to find and reconcile all the exceptions\footnote{We extend the
  notion of exceptions to denote all orbits that could fit a
  particular diagram but that were already counted in another
  diagram's contribution} and restrictions.  We believe that the
approach proposed in this paper helps to systematize the choice of
restrictions in such a way that we avoid treating exceptions
altogether.  This approach was inspired by the recent work on billiard
\cite{MHBHA05}.  It has its own shortcoming which will be discussed.
We believe that this shortcoming applies to \cite{MHBHA05} as well.

\section{A consistent approach to evaluating diagrams}
\label{sec:newmethod}

\subsection{The second order diagram}

\begin{figure}[t]
  \centerline{\includegraphics{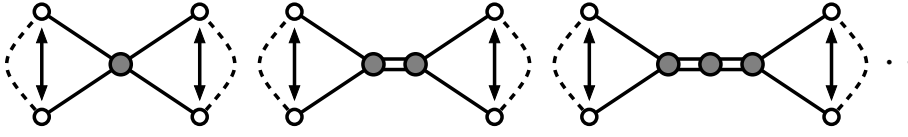}}
  \caption{The ``figure of eight'' orbits with intersection length 0,
    1 and 2.  To ensure that the length is well defined, we impose
    restrictions on both sides of the intersection.  The restrictions
    are drawn as double-ended arrows.}
  \label{fig:diag2}
\end{figure}

We start by illustrating our approach on the simplest nontrivial
diagram, the ``figure of eight''.  The trick is to introduce
restrictions on {\em both\/} sides of the ``extended'' intersection
and then sum over all possible intersection lengths, see
Fig.~\ref{fig:diag2}.

It turns out that we need to evaluate the first type of orbits on
Fig.~\ref{fig:diag2} separately.  The contribution of such orbits
reads
\begin{equation}
  \label{eq:diag2_0}
  K_{2,0} = \frac{n^2}{N^2} \sum_{v}\sum_{s_1,s_2,f_1,f_2}  
  \frac{C}{N^n} \e^{2\pi i(f_2-s_2)(f_1-s_1)/N}
  \left(1-\delta_{s_1,f_1}\right)
  \left(1-\delta_{s_2,f_2}\right),
\end{equation}
where we have implemented restrictions by using the Kronecker deltas.  The
term $C$ counts the number of ways to complete the orbit by choosing the arc
configurations.  The two arcs are uniquely defined by the vertices they pass
through.  Between the two of them, they pass through $n-6$ vertices (we have
already chosen 6 vertices of the orbit: $s_1$, $s_2$, $f_1$, $f_2$, and $v$
twice).  Thus there are $P(n)=n-5$ ways to choose the lengths of the arcs and
$N^{n-6}$ ways to choose the vertices themselves, leading to $C=(n-5)N^{n-6}$.
Going back to (\ref{eq:diag2_0}), we open one of the brackets to obtain
\begin{eqnarray}
  \label{eq:diag2_0_cont1}
  \nonumber 
  K_{2,0} &=& \frac{n^2}{N^2} \sum_{v}\sum_{s_1,s_2,f_1,f_2}  
  \frac{C}{N^n} \e^{2\pi i(f_2-s_2)(f_1-s_1)/N}
  \left(1-\delta_{s_2,f_2}\right) \\
  && - \frac{n^2}{N^2} \sum_{v}\sum_{s_1,s_2,f_1,f_2}  
  \frac{C}{N^n} \e^{2\pi i(f_2-s_2)(f_1-s_1)/N}
  \delta_{s_1,f_1} \left(1-\delta_{s_2,f_2}\right).
\end{eqnarray}
In both terms we perform the sum over $f_1$.  Since $\sum_{f_1}
\e^{2\pi i(f_2-s_2)f_1/N} = \delta_{s_2,f_2}$ and
$\delta_{s_2,f_2}\left(1-\delta_{s_2,f_2}\right) = 0$, the first sum
is identically 0.  In the second we substitute $f_1=s_1$ throughout to
get
\begin{eqnarray}
  \label{eq:diag2_0_cont2}
  \nonumber
  K_{2,0} &=& - \frac{n^2}{N^2} \sum_{v}\sum_{s_1,s_2,f_2}  
  \frac{C}{N^n} \left(1-\delta_{s_2,f_2}\right) = -
  \frac{n^2}{N^2} N^3(N-1) \frac{C}{N^n}\\
  &=& -\frac{n^2(n-5)(N-1)}{N^5}.
\end{eqnarray}
The obtained result is out of order: since $\tau=n/N^2$, we got an
answer of the order $\tau^3N^2$.  Thus it is vital to evaluate the
corrections from the extended intersections.

It is easy to see that for orbits with an extended intersection, $A_\p
A_\q^*$ is independent of the orbit configuration and is equal to
$N^{-n}$.  For an intersection of length $l_1$, there are
$N^{n-6-2l_1}(n-5-2l_1)$ possible arc configurations.  There are also
$N^{l_1+1}$ ways to choose the vertices denoted by the shaded circles
on Fig.~\ref{fig:diag2} and $N^2(N-1)^2$ ways to choose the vertices
denoted by the empty circles.  The total contribution is
\begin{equation*}
  \begin{split}
    K_{2,l_1} &= \frac{n^2}{N^2} N^{3+l_1}(N-1)^2
    \frac{N^{n-6-2l_1}(n-5-2l_1)}{N^n} \\
    &= \frac{n^2(n-5-2l_1)(N-1)^2}{N^{5+l_1}}.
  \end{split}
\end{equation*}
The contribution of all ``figure of eight'' orbits is thus
\begin{equation}
  \label{eq:diag2}
  \begin{split}
    K_2(\tau) &= -\frac{n^2(n-5)(N-1)}{N^5}  
    + \sum_{l_1=1}^\infty \frac{n^2(n-5-2l_1)(N-1)^2}{N^{5+l_1}}\\
    &= -2\frac{n^2}{N^4}.
  \end{split}
\end{equation}
Taking into account that $\tau=n/N^2$, we recover the correct
second-order correction $K_2(\tau) = -2\tau^2$.  If we restrict the
sum over $l_1$ to the values up to $(n-5)/2$, we get a correction term
of the form $2n^2/N^{(n+3)/2}$ which decays faster than exponentially
as we send $n$ and $N$ to infinity.

\subsection{Evaluation of some third order diagrams}

To gain some intuition for the general case we evaluate the
contributions of two of the five diagrams contributing to the third
order.  The reader is referred to \cite{BSW03} for a detailed list of
the third order diagrams and their ``multiplicities''.  The treatment
of these diagrams was also performed for disordered systems in
\cite{SLA98}, and for chaotic billiards in \cite{HMBH04}.

\begin{figure}[t]
  \centerline{\includegraphics{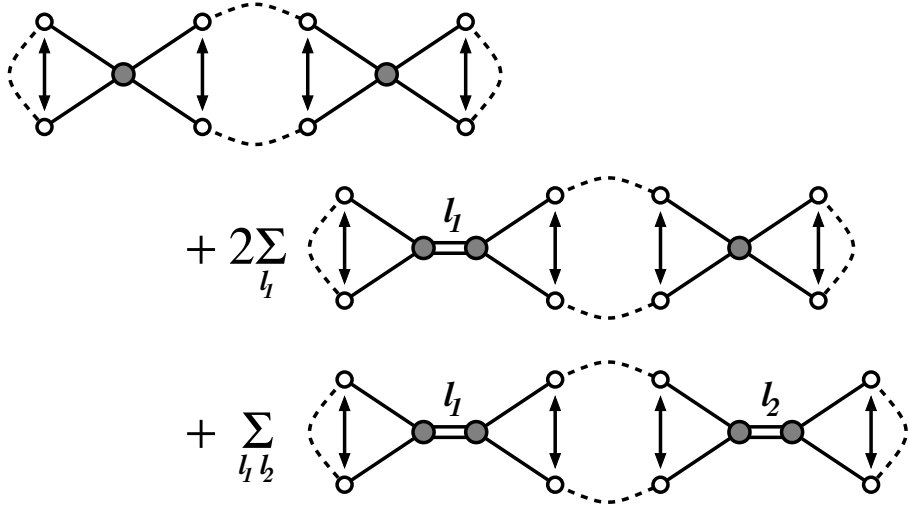}}
  \caption{The ``sausage'' orbits with various intersection lengths.
    The factor of 2 in the second line accounts for the symmetric case
    $l_2>0$, $l_1=0$.}
  \label{fig:diag3a}
\end{figure}

The first diagram we evaluate is the ``sausage'' orbits shown in
Fig.~\ref{fig:diag3a}.  Here we need to require that all arcs
(depicted by dashed lines) are at least one bond long.  As noted by
H.~Schanz \cite{schanz-private} this restriction excludes some
potentially contributing orbits.  Why they do not contribute to the
form factor is a non-trivial question.  In particular such orbits are
not ``rare'' enough to guarantee {\em a priori}\footnote{by that we
  mean an estimate based on the number of orbits and the modulus of an
  orbit pair's contribution, without taking into account the unitary
  cancellations} that their contribution is vanishingly small.  We
conducted a separate study to evaluate the missing orbits for this
particular diagram and discovered that cancellations akin to those in
(\ref{eq:diag2}) ensure they do not contribute.  However, we are not
aware of any general argument applicable to higher diagrams.

In the intersections of length 0 (first and second line in
Fig.~\ref{fig:diag3a}) we open up one of the restrictions, in the same
way as in (\ref{eq:diag2_0_cont1}).  Another component necessary
is the number of ways to complete the orbit by choosing the
configurations of the 4 arcs.  This is equal to
$P(n,l_1,l_2)N^{n-12-2l_1-2l_2}$, where 
\begin{equation}
  \label{eq:arcs}
  P(n, l_1, l_2) = \binom{n-9-2l_1-2l_2}{3}
\end{equation}
is the number of possible choices for the lengths of the arcs --- essentially
the number of partitions of $n$ into an ordered sum of 4 natural numbers.
Now the total contribution of the diagram is
\begin{multline}
  \label{eq:diag3a}
  K_{3a} = \frac{n^2}{N^2} P(n,0,0) \frac{(N-1)^2}{N^6} 
  - 2 \frac{n^2}{N^2} \sum_{l_1=1}^\infty P(n, l_1, 0) 
  \frac{(N-1)^3}{N^{6+l_1}} \\
  + \frac{n^2}{N^2} \sum_{l_1,l_2=1}^\infty P(n,l_1,l_2) 
  \frac{(N-1)^4}{N^{6+l_1+l_2}} \\
  = 4\frac{n^3}{N^6} - \frac{n^2(48N-32)}{N^6(N-1)} 
  = 4\tau^3\left(1+O(n^{-1})\right).
\end{multline}

\begin{figure}[t]
  \centerline{\includegraphics{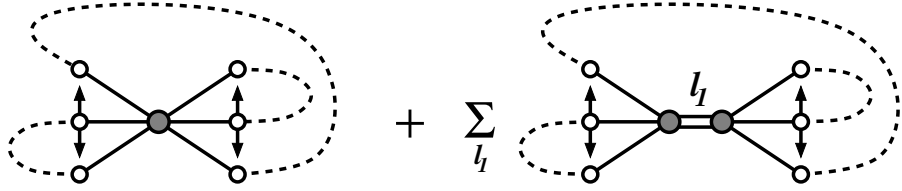}}
  \caption{One of the ``clover leaf'' diagrams contributing to the
  third order.  In a triple intersection, such as above, a
  restriction prohibits coincidence of {\em all\/} 3 vertices at the
  same time.}
  \label{fig:diag3b}
\end{figure}

Next we evaluate a diagram with a triple intersection, shown in
Fig.~\ref{fig:diag3b}.  The restrictions now mean that the three
vertices cannot be all equal.  In terms of Kronecker deltas it is
implemented as $\left(1-\delta_{v_1,v_2}\delta_{v_1,v_3}\right)$.  We
again require that all arcs have non-zero bond lengths; this
requirement excludes some orbits that ought to be counted.  We must
also note that in this case the restrictions do not fully rule out
counting of orbits that ought to be excepted, for example the orbits
with self-retracing arcs.  The contributions of such orbits cannot be
ruled out {\em a priori\/} --- they occur often enough to contribute to the
third order in $\tau$.  However they do not contribute (as signified
by the results of \cites{BSW03,B04}) due to cancellations which
should be studied separately.

Turning a blind eye to these special orbits, we can evaluate the
contributions of Fig.~\ref{fig:diag3b}.  There are now $P(n,l_1)N^{n-9-3l_1}$
arc configurations, where
\begin{equation}
  \label{eq:arcs_diag3b}
  P(n,l_1) = \binom{n-7-3l_1}{2}.
\end{equation}
Dealing with $l_1=0$ case separately we arrive to
\begin{equation}
  \label{eq:diag3b}
  \begin{split}
    K_{3b} &= - \frac{n^2}{N^2} P(n,0)\frac{(N^2-1)}{N^6}
    + \frac{n^2}{N^2} \sum_{l_1=1}^\infty P(n,l_1)
    \frac{(N^2-1)^2}{N^{6+2l_1}} \\
    &= -3\frac{n^3}{N^6} + \frac{n^2(27N^2-18)}{N^6(N^2-1)} 
    = -3\tau^3\left(1+O(n^{-1})\right).
  \end{split}
\end{equation}

\subsection{The general formula}

The big advantage of the new way of evaluating diagrams is that the
restrictions we impose essentially decouple different intersections
from each other.  This, in turn, allows us to write a general formula
for a contribution of a diagram using only the information about the
number of intersections and not how they are interconnected.

Consider the ``sausage'' diagram again.  To write its contribution in
a more uniform fashion we introduce a special factor, defined for
$r\geq a\geq0$,
\begin{equation}
  \label{eq:M_factor}
  M_a^r = \left\{
    \begin{array}{ll}
      1 &\mbox{ if } a=0 \mbox{ and } r=0,\\
      0 &\mbox{ if } a=0 \mbox{ and } r\neq0,\\
      \binom{r-1}{a-1} & \mbox{ otherwise.}
    \end{array}
  \right.
\end{equation}
We can now rewrite the sums for $K_{3a}$ in the following form:
\begin{multline}
  \label{eq:diag3a_re}
  K_{3a} = \frac{n^2}{N^2}\sum_{r=0}^\infty P(n,r) M_0^r 
  \frac{(N-1)^2}{N^{6+r}} \\
  - 2 \frac{n^2}{N^2} \sum_{r=1}^\infty P(n,r) M_1^r
  \frac{(N-1)^3}{N^{7+r}}
  + \frac{n^2}{N^2} \sum_{r=2}^\infty P(n,r) M_2^r
  \frac{(N-1)^4}{N^{8+r}},
\end{multline}
where
\begin{equation*}
  P(n,r) = \binom{n-9-2r}{3}.
\end{equation*}
In the first summand of (\ref{eq:diag3a_re}), $r$ is a dummy
variable since $M_0^r=0$ unless $r=0$.  In the second summand $r$
corresponds to $l_1$ in the second term of (\ref{eq:diag3a}).
Finally, in the third summand $r=l_1+l_2$ and $M_2^r$ counts the number
of ways $r$ can be represented in such a way.  Now we can generalize
further by denoting by $v_2$ the number of 2-intersections in the
diagram (in our case $v_2=2$) and by introducing the notation $A=4v_2+1+2r$:
\begin{eqnarray}
  \label{eq:2inter}
  K_{3a} &=& \frac{n^2}{N^2} \sum_{a=0}^{v_2} (-1)^{{v_2}-a} \binom{v_2}{a}
  \sum_{r=a}^\infty \binom{n-A}{2v_2-1} M_a^r
  \frac{(N-1)^{v_2+a}}{N^{r+3v_2}}.
\end{eqnarray}

Using the same transformation, the contribution $K_{3b}$ can be written
as
\begin{equation}
  \label{eq:diag2b_re}
  K_{3b}
  = \frac{n^2}{N^2} \sum_{a=0}^{v_3} (-1)^{{v_3}-a} \binom{v_3}{a}
  \sum_{r=a}^\infty \binom{n-A}{3v_3-1} M_a(r)
  \frac{(N^2-1)^{v_3+a}}{N^{6v_3+r}},
\end{equation}
where $v_3=1$ and $A = 6v_3+3r+1$.  From here we generalize in the
following fashion.  Let, for a given diagram, the vector $\vv =
(v_2,v_3,\ldots)$ describe the number of intersections of
multiplicities $2$, $3$ etc.  Let $k$ be the maximal multiplicity in
the diagram: $v_j=0$ for $j>k$.  Then the contribution of this diagram
is given by
\begin{multline}
  \label{eq:gen_contr}
  D(\vv) = \frac{n^2}{N^2} \sum_{a_2=0}^{v_2} \cdots \sum_{a_k=0}^{v_k}
  \prod_{i=2}^k (-1)^{v_i-a_i} \binom{v_i}{a_i} \\
    \times \left( \sum_{r_2=a_2}^\infty \cdots \sum_{r_k=a_k}^\infty
    \binom{n-A}{L-1} N^{-P} 
    \prod_{i=2}^k M_{a_i}(r_i) \left(N^{i-1}-1\right)^{v_i+a_i} \right),
\end{multline}
where 
\begin{equation*}
  \label{eq:defn_A}
  A = \sum_{i=2}^k i(2v_i + r_i) + 1,\quad
  L = \sum_{i=2}^k iv_i, \quad
  P = \sum_{i=2}^k (i-1)(3v_i + r_i).
\end{equation*}

We studied sum (\ref{eq:gen_contr}) for several values of $(v_2, v_3,
\ldots)$ and found that it gives the contribution derived in
\cite{MHBHA05},
\begin{equation}
  \label{eq:gen_cont_res}
  D(\vv) = (-1)^V \prod_{i=2}^k i^{v_i} \frac{\tau^{L-V+1}}{(L-V-1)!}
  \left(1+O(n^{-1})\right),
\end{equation}
where
\begin{equation*}
  Q = \prod_{i=2}^k i^{v_i} \quad \mbox{and} \quad
  V = \sum_{i=2}^k v_i.
\end{equation*}
At this point we can evoke the combinatorial results of \cite{MHBHA05}
for the number of diagrams with the characteristic $\vv$.  The
summation of the leading order term of (\ref{eq:gen_cont_res}) over
all possible diagrams then reproduces the predicted Random Matrix
result, (\ref{eq:K_GOE}).

\section{Conclusions, comparisons with \cite{MHBHA05} and the outlook}
\label{sec:conclusion}

In the present article we outline a construction which allows us to
recover expression (\ref{eq:gen_cont_res}) for the contribution of a
general diagram to the form factor of a complete Fourier quantum
graph.  The summation over all diagrams, described in \cite{MHBHA05}
would then recover the predicted Random Matrix result.  However the
final result, (\ref{eq:gen_contr}), was obtained at the cost of
ignoring the orbits which can be generally characterized as having
short stretches between self-intersections.  We believe that the same
omission was used in \cite{MHBHA05}, described in particular in the
paragraph immediately following Eq.(16) (the discussion in Appendix
D.2 is also relevant).  While not claiming that the same applies to
billiard systems, we can testify that on graphs the contribution of
the orbits with very short stretches is {\em absolutely divergent} in
the limit $B\to\infty$.  This underlines the need to analyze the
unitary cancellations which ensure that such orbits do not contribute
to the final answer.  The missing analysis also holds the key to the
possibility of extending the discussed summation to systems with
slower mixing.

Now that we have more understanding of how the universal contributions to the
form factor arise from the sum of periodic orbits in the region $\tau<1$, the
question of why the expansion techniques break down beyond the Heisenberg time
$\tau=1$ comes to the forefront.  It seems, that if we are to recover the
Random Matrix prediction for $\tau>1$, the periodic orbit contributions would
have to be classified in an entirely different way.  Quite possibly, the
summation needs to be based on the degeneracy classes --- the sets of orbits
of the same length --- rather than on the diagram partition which is in a way
transversal to the degeneracy class partition.  Such degeneracy-based
summation has been performed before for a special type of graphs, the Neumann
star graphs \cite{BK99} (see also \cite{keating_here} for an overview of the
results).  The radius of convergence of the expansion was found to be finite
but, if re-summed \cite{BBK01}, the result was valid for all values of $\tau$.
Important input for the question of going beyond $\tau=1$ can also be obtained
by comparing the periodic orbit expansions with the non-perturbative methods
of \cites{GA04,GA05} which are valid for all $\tau$.


\end{document}